\input harvmac.tex

\vskip 1.5in
\Title{\vbox{\baselineskip12pt
\hbox to \hsize{\hfill}
\hbox to \hsize{\hfill WITS-CTP-059}}}
{\vbox{
	\centerline{\hbox{Higher Spins and Open Strings:
		}}\vskip 5pt
        \centerline{\hbox{Quartic Interactions
		}} } }
\centerline{Dimitri Polyakov\footnote{$^\dagger$}
{dimitri.polyakov@wits.ac.za}}
\medskip
\centerline{\it National Institute for Theoretical Physics (NITHeP)}
\centerline{\it  and School of Physics}
\centerline{\it University of the Witwatersrand}
\centerline{\it WITS 2050 Johannesburg, South Africa}
\vskip .3in

\centerline {\bf Abstract}
We analyze quartic gauge-invariant interactions
of massless higher spin fields by using the 
vertex operators constructed in our previous  works
and computing their four-point amplitudes in superstring theory.
The behaviour of the amplitudes is quite different
from the standard Veneziano structure, due to their
nonstandard ghost coupling.
The kinematic
 part of the quartic interactions of the higher spins is determined 
by the matter structure of their vertex operators while
 nonlocality of the interactions is the consequence of 
the ghost structure of these operators.
We compute explicitly  the four-point  amplitude
describing the complete gauge-invariant $1-1-3-3$ quartic interaction
(two massless spin 3  particles interacting with two photons)
and comment on more general  $1-1-s-s$ cases, particularly
pointing out the structure of $1-1-5-5$ coupling.

\Date{November 2010}
\vfill\eject
\lref\bianchi{ M. Bianchi, V. Didenko, arXiv:hep-th/0502220}
\lref\sagnottia{A. Sagnotti, E. Sezgin, P. Sundell, hep-th/0501156}
\lref\sorokin{D. Sorokin, AIP Conf. Proc. 767, 172 (2005)}
\lref\fronsdal{C. Fronsdal, Phys. Rev. D18 (1978) 3624}
\lref\coleman{ S. Coleman, J. Mandula, Phys. Rev. 159 (1967) 1251}
\lref\haag{R. Haag, J. Lopuszanski, M. Sohnius, Nucl. Phys B88 (1975)
257}
\lref\weinberg{ S. Weinberg, Phys. Rev. 133(1964) B1049}
\lref\fradkin{E. Fradkin, M. Vasiliev, Phys. Lett. B189 (1987) 89}
\lref\skvortsov{E. Skvortsov, M. Vasiliev, Nucl.Phys.B756:117-147 (2006)}
\lref\skvortsovb{E. Skvortsov, J.Phys.A42:385401 (2009)}
\lref\mva{M. Vasiliev, Phys. Lett. B243 (1990) 378}
\lref\mvb{M. Vasiliev, Int. J. Mod. Phys. D5
(1996) 763}
\lref\mvc{M. Vasiliev, Phys. Lett. B567 (2003) 139}
\lref\brink{A. Bengtsson, I. Bengtsson, L. Brink, Nucl. Phys. B227
 (1983) 31}
\lref\deser{S. Deser, Z. Yang, Class. Quant. Grav 7 (1990) 1491}
\lref\bengt{ A. Bengtsson, I. Bengtsson, N. Linden,
Class. Quant. Grav. 4 (1987) 1333}
\lref\boulanger{ X. Bekaert, N. Boulanger, S. Cnockaert,
J. Math. Phys 46 (2005) 012303}
\lref\metsaev{ R. Metsaev, arXiv:0712.3526}
\lref\siegel{ W. Siegel, B. Zwiebach, Nucl. Phys. B282 (1987) 125}
\lref\siegelb{W. Siegel, Nucl. Phys. B 263 (1986) 93}
\lref\nicolai{ A. Neveu, H. Nicolai, P. West, Nucl. Phys. B264 (1986) 573}
\lref\damour{T. Damour, S. Deser, Ann. Poincare Phys. Theor. 47 (1987) 277}
\lref\sagnottib{D. Francia, A. Sagnotti, Phys. Lett. B53 (2002) 303}
\lref\sagnottic{D. Francia, A. Sagnotti, Class. Quant. Grav.
 20 (2003) S473}
\lref\sagnottid{D. Francia, J. Mourad, A. Sagnotti, Nucl. Phys. B773
(2007) 203}
\lref\labastidaa{ J. Labastida, Nucl. Phys. B322 (1989)}
\lref\labastidab{ J. Labastida, Phys. rev. Lett. 58 (1987) 632}
\lref\mvd{L. Brink, R.Metsaev, M. Vasiliev, Nucl. Phys. B 586 (2000)183}
\lref\klebanov{ I. Klebanov, A. M. Polyakov,
Phys.Lett.B550 (2002) 213-219}
\lref\mve{
X. Bekaert, S. Cnockaert, C. Iazeolla,
M.A. Vasiliev,  IHES-P-04-47, ULB-TH-04-26, ROM2F-04-29, 
FIAN-TD-17-04, Sep 2005 86pp.}
\lref\sagnottie{A. Campoleoni, D. Francia, J. Mourad, A.
 Sagnotti, Nucl. Phys. B815 (2009) 289-367}
\lref\sagnottif{
A. Campoleoni, D. Francia, J. Mourad, A.
 Sagnotti, arXiv:0904.4447}
\lref\sagnottig{D. Francia, A. Sagnotti, J.Phys.Conf.Ser.33:57 (2006)}
\lref\selfa{D. Polyakov, Int.J.Mod.Phys.A20:4001-4020,2005}
\lref\selfb{ D. Polyakov, arXiv:0905.4858}
\lref\selfc{D. Polyakov, arXiv:0906.3663}
\lref\selfd{D. Polyakov, Phys.Rev.D65:084041 (2002)}
\lref\mirian{A. Fotopoulos, M. Tsulaia, Phys.Rev.D76:025014,2007}
\lref\extraa{I. Buchbinder, V. Krykhtin,  arXiv:0707.2181}
\lref\extrab{I. Buchbinder, V. Krykhtin, Phys.Lett.B656:253-264,2007}
\lref\extrac{X. Bekaert, I. Buchbinder, A. Pashnev, M. Tsulaia,
Class.Quant.Grav. 21 (2004) S1457-1464}
\lref \extrad{I. Buchbinder, A. Pashnev, M. Tsulaia,
arXiv:hep-th/0109067}
\lref\extraf{I. Buchbinder, A. Pashnev, M. Tsulaia, 
Phys.Lett.B523:338-346,2001}
\lref\extrag{I. Buchbinder, E. Fradkin, S. Lyakhovich, V. Pershin,
Phys.Lett. B304 (1993) 239-248}
\lref\extrah{I. Buchbinder, A. Fotopoulos, A. Petkou, 
 Phys.Rev.D74:105018,2006}
\lref\bonellia{G. Bonelli, Nucl.Phys.B {669} (2003) 159}
\lref\bonellib{G. Bonelli, JHEP 0311 (2003) 028}
\lref\ouva{C. Aulakh, I. Koh, S. Ouvry, Phys. Lett. 173B (1986) 284}
\lref\ouvab{S. Ouvry, J. Stern, Phys. Lett.  177B (1986) 335}
\lref\ouvac{I. Koh, S. Ouvry, Phys. Lett.  179B (1986) 115 }
\lref\hsself{D.Polyakov, arXiv:1005.5512}
\lref\sundborg{ B. Sundborg, ucl.Phys.Proc.Suppl. 102 (2001)}
\lref\sezgin{E. Sezgin and P. Sundell,
Nucl.Phys.B644:303- 370,2002}
\lref\morales{M. Bianchi,
J.F. Morales and H. Samtleben, JHEP 0307 (2003) 062}
\lref\giombif{S. Giombi, Xi Yin, arXiv:0912.5105}
\lref\giombis{S. Giombi, Xi Yin, arXiv:1004.3736}
\lref\bekaert{X. Bekaert, N. Boulanger, P. Sundell, arXiv:1007.0435}
\lref\taronna{A. Sagnotti, M. Taronna, arXiv:1006.5242, 
Nucl.Phys.B842:299-361,2011}
\lref\zinoviev{Yu. Zinoviev, arXiv:1007.0158}
\lref\fotopoulos{A. Fotopoulos, M. Tsulaia, arXiv:1007.0747}
\lref\fotopouloss{A. Fotopoulos, M. Tsulaia, arXiv:1009.0727}
\lref\taronnao{M. Taronna, arXiv:1005.3061}
\lref\per{	
N. Boulanger,S. Leclercq, P. Sundell, JHEP 0808(2008) 056 }
\lref\mav{V. E. Lopatin and M. A. Vasiliev, Mod. Phys. Lett. A 3 (1988) 257}
\lref\zinov{Yu. Zinoviev, Nucl. Phys. B 808 (2009)}
\lref\sv{E.D. Skvortsov, M.A. Vasiliev,
Nucl. Phys.B 756 (2006)117}
\lref\mvasiliev{D.S. Ponomarev, M.A. Vasiliev, Nucl.Phys.B839:466-498,2010}
\lref\zhenya{E.D. Skvortsov, Yu.M. Zinoviev, arXiv:1007.4944}
\lref\perf{N. Boulanger, C. Iazeolla, P. Sundell, JHEP 0907 (2009) 013 }
\lref\pers{N. Boulanger, C. Iazeolla, P. Sundell, JHEP 0907 (2009) 014 }
\lref\selft{D. Polyakov,Phys.Rev.D82:066005,2010}
\lref\selftt{D. Polyakov, Int.J.Mod.Phys.A25:4623-4640,2010}
\lref\tseytlin{I. Klebanov, A Tseytlin, Nucl.Phys.B546:155-181,1999}
\lref\ruben{R. Manvelyan, K. Mkrtchyan, W. Ruehl, arXiv:1009.1054}
\lref\rubenf{R. Manvelyan, K. Mkrtchyan, W. Ruehl, Nucl.Phys.B836:204-221,2010}
\lref\robert{
R. De Mello Koch, A. Jevicki, K. Jin, J. A. P. Rodrigues, arXiv:1008.0633}
\centerline{\bf  1. Introduction}

Constructing consistent gauge-invariant 
field theories of interacting higher spins
is an important and fascinating problem that has attracted 
deep interest. Despite many efforts
by the leading experts in the field and some remarkable results
over recent years
(for an incomplete list of references, see
~{\fronsdal, \bianchi, \sagnottia, \sagnottib, \sagnottic,
\sagnottid, \sorokin, 
\mva, \mvb, \mvc, \mvasiliev, \deser, \bengt, \siegel, \siegelb,
\nicolai, \damour, \brink, \boulanger,
\labastidaa, \labastidab, \mvd, \mve, \mvasiliev, \mirian, \extrah,
\bonellia,\extrad, \bekaert, 
\perf, \pers, \taronna, \taronnao, 
\zinoviev,
\fotopoulos, \fotopouloss, \taronna,\zinoviev, \bekaert, \morales, \giombif, 
\giombis,
\sezgin, \sundborg, \per, \zhenya, \robert, \ruben, \rubenf})

the entire subject is well known to be difficult to approach.
In particular, while there has been some progress in formulating
free higher spin field theories as well as those with
cubic interactions, our understanding of higher
order interactions (such as quartic) is still
very limited.

There are many reasons why the field theories of spins greater than 2
are of interest and importance. To mention some of them,
while it may not seem plausible
 that higher spin particles could ever be observed
in four-dimensional world, objects such as higher spins are likely
to be present in higher dimensional physics . Higher spin 
fields in
AdS space are known to be important ingredient in AdS/CFT correspondence
~{\klebanov};
in addition, constructing gauge invariant interactions of higher
spins is by itself an interesting and challenging mathematical problem.
 String theory appears to be a particularly efficient
framework to approach the problem of higher spins.  One reason for this
is that the vertex operators describing the emissions of higher
spins by a string, appear very naturally in the massive sector of 
string theory (although the mass to spin relations for such operators
are usually quite rigid, with $m^2$ roughly proportional to the
spin value $s$.)
 One could then consider the tensionless limit $\alpha^{\prime}\rightarrow
\infty$ in which the higher spin operators formally become massless.
There are several difficulties one faces in this approach.
Firstly, the space-time fields coupling to the massive operators
usually would lack the gauge symmetries necessary to ensure the consistency
of the interactions, and it is not clear how to recover these symmetries
in the tensionless limit. Secondly, to recover the
 gauge-invariant interactions of the higher spins from string theory
correlators, one generally has to consider the low energy limit
of string theory,
which of course is different from the tensionless limit.
 In our previous works ~{\selft, \selftt} we have constructed the open string
vertex operators that describe the higher spin fields
with spin values from 3 to 9,
which are massless at an arbitrary tension due to their 
nontrivial couplings to the $\beta-\gamma$ ghost system.
The explicit expressions for these vertex operators are given by:

\eqn\grav{\eqalign{
V_{s=3}(p)=H_{a_1a_2a_3}(p)c{e}^{-3\phi}\partial{X^{a_1}}\partial{X^{a_2}}
\psi^{a_3}e^{i{\vec{p}}{\vec{X}}}
\cr
V_{s=4}(p)=H_{a_1...a_4}(p)c\eta{e^{-4\phi}}\partial{X^{a_1}}\partial{X^{a_2}}
\partial\psi^{a_3}\psi^{a_4}e^{i{\vec{p}}{\vec{X}}}\cr
V_{s=5}(p)=H_{a_1...a_5}(p)c{e^{-4\phi}}\partial{X^{a_1}}...\partial{X^{a_3}}
\partial\psi^{a_4}\psi^{a_5}e^{i{\vec{p}}{\vec{X}}}\cr
V_{s=6}(p)
=H_{a_1...a_6}(p)c\eta{e^{-5\phi}}\partial{X^{a_1}}...\partial{X^{a_3}}
\partial^2\psi^{a_4}\partial\psi^{a_5}\psi^{a_6}e^{i{\vec{p}}{\vec{X}}}\cr
V_{s=7}(p)=H_{a_1...a_7}(p)c{e^{-5\phi}}\partial{X^{a_1}}...\partial{X^{a_4}}
\partial^2\psi^{a_5}\partial\psi^{a_6}\psi^{a_7}e^{i{\vec{p}}{\vec{X}}}\cr
V_{s=8}(p)=
H_{a_1...a_8}(p)c\eta{e^{-5\phi}}\partial{X^{a_1}}...\partial{X^{a_7}}
\psi^{a_8}e^{i{\vec{p}}{\vec{X}}}\cr
V_{s=9}(p)=H_{a_1...a_9}(p)c{e^{-5\phi}}\partial{X^{a_1}}...\partial{X^{a_8}}
\psi^{a_9}e^{i{\vec{p}}{\vec{X}}}
}}

where $X^{a}$ and $\psi^a$ are the RNS worldsheet bosons
 and
fermions ($a=0,...,d-1$), 
the ghost fields are bosonized according to

\eqn\grav{\eqalign{b=e^{-\sigma},c=e^{\sigma}\cr
\gamma=e^{\phi-\chi}\equiv{e^\phi}\eta\cr
\beta=e^{\chi-\phi}\partial\chi\equiv\partial\xi{e^{-\phi}}}}
The operators (1) are picture inequivalent
and are the elements of ghost cohomologies
$H_{-3}$,$H_{-4}$ and $H_{-5}$. All the expressions
for the operators (1) are given at their minimal negative
superconformal ghost pictures (e.g. $-3$ for $s=3$ and $-5$ for $s=9$)
at which they are annihilated by the direct picture-changing
transformation.
The  symmetric tensors $H_{a_1...a_s}(p)$ describe massless
higher spin fields in space-time, with the spin values
$3\leq{s}\leq{9}$.
The equations of motion and the gauge symmetry transformations
follow from the BRST constraints on the operators (1) ~{\selft}.
Namely,
the on-shell Fierz-Pauli constraints:
\eqn\grav{\eqalign{
H^{a_1}_{a_1a_3...a_S}(p)=0\cr
p^{a_1}H_{a_1...a_S}(p)=0\cr
p^2H_{a_1...a_s}(p)=0}}
follow from the invariance condition
$\lbrace{Q},V_s\rbrace=0$
where

\eqn\lowen{Q_=Q_1+Q_2+Q_3}
is the BRST operator with

\eqn\grav{\eqalign{
Q_1=\oint{{dz}\over{2i\pi}}\lbrace{cT-bc\partial{c}}\rbrace\cr
Q_2=-{1\over2}\oint{{dz}\over{2i\pi}}\gamma\psi_a\partial{X^a}\cr
Q_3=-{1\over4}\oint{{dz}\over{2i\pi}}b\gamma^2}}

The BRST nontriviality conditions, in turn, entail the gauge symmetry 
transformations for the higher spins ~{\selftt}.
For the symmetric tensors, the transformations are given by
\eqn\grav{\eqalign{
H_{a_1...a_s}(p)\rightarrow{H_{a_1...a_s}}(p)+
p_{(a_1}\Lambda_{a_2...a_s)}(p)}}
(where $\Lambda$ is also traceless and symmetric)
as under the
the shift of symmetric H-tensors by symmetrized derivatives
of $\Lambda$ the vertex operators (1) are shifted by the terms not contributing 
to correlation functions.
Therefore the gauge invariance
of the interaction terms for the higher spins, obtained
 in the field theory limit of string theory, is ensured by construction,
since the structure of these terms is entirely determined by the correlation
functions in string theory.
For detailed BRST analysis of the operators (1) see ~{\selft};
below, we shall briefly review the relation between
BRST constraints, equations of motion and gauge symmetries on 
the example of
 the $s=3$ operator (the $s>3$ cases are treated similarly).
The vertex operator for $s=3$ is given by:
\eqn\grav{\eqalign{
V_{s=3}(p)=H_{abc}(p)c{e}^{-3\phi}\partial{X^{a}}\partial{X^{b}}
\psi^{c}e^{i{\vec{p}}{\vec{X}}}
}}
This operator commutes with $Q_2$ and $Q_3$ of the BRST charge.
To commute with $Q_1$ it has to be a dimension 0 primary, i.e.
its OPE with stress-energy tensor must not contain singularities
stronger than a simple pole. This entails constraints on the 
rank 3 $H$-tensor. For general $H$, the OPE contains singularities
up to quartic pole, so  to ensure the commutation with $Q_1$
the coefficients in front of quartic, triple and double poles
must vanish separately. This leads to tracelessness, transversality
and masslessness conditions respectively, i.e. to the Fierz-Pauli
constraints (3) on $H$.
At the same time, the shift (6) shifts the operator (7) by terms
not contributing to correlation functions.
To see this, consider the general (not necessarily symmetric)
tensor $H_{a|bc}$ (note that the form of constraints
(3) following from BRST-invariance arguments does not
depend on the symmetric propertirs of $H$ and remains the same). 
Under the shift $H_{a|bc}(p)\rightarrow{H_{a|bc}}(p)+p_c\Lambda_{ab}(p)$
where $\Lambda$ is symmetric and traceless, the operator (7)
is shifted by the BRST-exact part
\eqn\grav{\eqalign{{\sim}ce^{-3\phi}({\vec{p}}{\vec{\psi}})\Lambda_{ab}
\partial{X^a}\partial{X^b}e^{i{\vec{p}}{\vec{X}}}\sim
\cr
\lbrace{Q},ce^{\chi-4\phi}\partial\chi({\vec{p}}{\vec{\psi}})
({\vec{\psi}}\partial{\vec{X}})\Lambda_{ab}\partial{X^a}\partial{X^b}
e^{i{\vec{p}}{\vec{X}}}\rbrace}}
which insertion to any correlator is zero.
On the other hand, the tensor $p_c\Lambda_{ab}$ can be decomposed as
\eqn\grav{\eqalign{p_c\Lambda_{ab}={1\over2}(p_{(c}\Lambda_{ab)}
+p_{\lbrack{c}}\Lambda_{a\rbrack{b}})}}

and insertions of operators corresponding to different Young tableau 
to correlation functions  vanish
separately.
As a matter of fact, vanishing of
$p_{\lbrack{c}}\Lambda_{a\rbrack{b}}$-type
insertions to correlators is 
a just a particular example of a general property of
 $S$-matrix elements of $s=3$
vertex operators coupling $3$-tensors with hook-like
Young diagrams - it can be shown that such operators do not
contribute to S-matrices, which is reminiscent
of what happens in the frame-like approach ~{\mav, \zinov, \sv, \zhenya,
\mvasiliev}
 where contributions with hook-like symmetries are
eliminated by algebraic constraints.

Therefore the correlators are invariant under
shifting symmetric tensor $H_{abc}$ by symmetrized derivative
of $\Lambda$ , implying the gauge symmetry (6) in the field theory limit.
In order to compute the correlation functions involving the operators (1),
one also needs their representations in dual positive ghost pictures.
In order to obtain the positive picture presentation for  elements
of $H_{-n-2}$ ( physical operators existing at minimal negative
picture $-n-2$ and below;$n=1,2,...$)
one has to replace $e^{-(n+2)\phi}$ with $e^{n\phi}$ (without
changing the matter part) and perform the homotopy transformation
using the $K$-operator ~{\selfc}.
Namely, if a higher spin vertex at minimal negative picture
$-n-2$ has the structure
\eqn\lowen{V_{-n-2}=ce^{-(n+2)\phi}F_{{{n^2}\over2}+n+1}(X,\psi)}
where $F_{{{n^2}\over2}+n+1}(X,\psi)$ the is matter
primary field of conformal dimension ${{n^2}\over2}+n+1$,
one starts with the operator
\eqn\lowen{\oint{V_n}\equiv
\oint{{dz}}e^{n\phi}F_{{{n^2}\over2}+n+1}(X,\psi)}
This charge commutes with $Q_1$ since it is a worldsheet integral
of dimension 1 and $b-c$ ghost number zero but doesn't commute with
$Q_2$ and $Q_3$. To make it BRST-invariant, one has to
add the correction terms by using the following procedure 
~{\selfb, \selfc}.
We write
\eqn\grav{\eqalign{\lbrack{Q_{brst}},V_n(z)\rbrack=\partial{U}(z)+W_1(z)
+W_2(z)}}
and therefore
\eqn\lowen{\lbrack{Q_{brst}},
\oint{dz}V_n{\rbrack}=\oint{{dz}}(W_1(z)+W_2(z))}
where 
\eqn\grav{\eqalign{U(z)\equiv{cV_n(z)}\cr
\lbrack{Q_1,V_n}\rbrack=\partial{U}\cr
W_1=\lbrack{Q_2,V_n}\rbrack\cr
W_2=\lbrack{Q_3},V_n\rbrack}}
Introduce the dimension 0 $K$-operator:
\eqn\lowen{K(z)=-4c{e}^{2\chi-2\phi}(z)\equiv{\xi}\Gamma^{-1}(z)}
satisfying
\eqn\lowen{\lbrace{Q_{brst}},K\rbrace=1}
It is easy to check that this operator has a non-singular
operator product with $W_1$:
\eqn\lowen{K(z_1)W_1(z_2)\sim{(z_1-z_2)^{2n}}Y(z_2)+O((z_1-z_2)^{2n+1})}
where $Y$ is some operator of dimension $2n+1$.
Then the complete BRST-invariant operator
can be obtained from $\oint{dz}V_n(z)$
by the following transformation:

\eqn\grav{\eqalign{
\oint{dz}{V_n}(z){\rightarrow}A_n(w)=\oint{dz}V_n(z)+{{1}\over{(2n)!}}
\oint{dz}(z-w)^{2n}:K\partial^{2n}{(W_1+W_2)}:(z)
\cr
+{1\over{{(2n)!}}}\oint{{dz}}\partial_z^{2n+1}{\lbrack}
(z-w)^{2n}{K}(z)\rbrack{K}\lbrace{Q_{brst}},U\rbrace}}
where $w$ is some arbitrary point on the worldsheet.
It is then straightforward to check the invariance
of $A_n$ by using some partial integration along with
the relation (34) as well as the obvious identity
\eqn\lowen{\lbrace{Q_{brst}},W_1(z)+W_2(z)\rbrace=
-\partial(\lbrace{Q_{brst}},U(z)\rbrace)}
Although the invariant operators $A_n(w)$ depend on an
arbitrary point $w$ on the worldsheet, this dependence
is irrelevant in the correlators
 since all the $w$ derivatives  of $A_n$ are BRST exact -
the triviality of the derivatives ensures that
 there will be no $w$-dependence in any correlation
functions involving $A_n$.
Alternative (yet technically more complicated) method 
to obtain the positive picture representations for the higher spin
operators 
 is to use 
sequences of $Z$-transformations combined with picture changing
~{\selfc}
Namely, introduce 
the $Z$-operator, transforming the $b-c$ pictures (in particular,
mapping integrated vertices to unintegrated)
 given by ~{\selfa}
\eqn\lowen{Z(w)=b\delta(T)(w)=
\oint{dz}(z-w)^3(bT+4c\partial\xi\xi{e^{-2\phi}}T^2)(z)}
where $T$ is the full stress-energy tensor in RNS theory.
The usual picture-changing operator,
transforming the $\beta-\gamma$ ghost pictures, is given by
 $\Gamma(w)=:\delta(\beta)G:(w)
=:e^\phi{G}:(w)$.
Introduce the  $integrated$ picture-changing operators
$R_n(w)$ according to
\eqn\grav{\eqalign{R_{n}(w)=Z(w):\Gamma^{n}:(w)}}
where $:\Gamma^n:$ is the $n$th power of the standard
picture-changing operator:
\eqn\grav{\eqalign{
:\Gamma^{n}:(w)=:e^{{n}\phi}\partial^{n-1}G...\partial{G}G:(w)\cr
\equiv:\partial^{n-1}\delta(\beta)...\partial\delta(\beta)\delta(\beta):}}
Then the positive picture representations for the higher
spin operators $A_n$ can be obtained from the negative ones $V_{-n-2}$ (1)
by the transformation:
\eqn\lowen{A_n(w)=(R_2)^{n+1}(w){V_{-n-2}}(w)}

Since  both $Z$ and $\Gamma$ are BRST-invariant and nontrivial,
the $A_n$-operators by construction 
satisfy the BRST-invariance and non-triviality
conditions identical to those satisfied by their negative picture
counterparts
$V_{-2n-2}$ and therefore lead to the same Pauli-Fierz on-shell
conditions (3) and the gauge symmetries (6) for the higher spin fields.
For the $s=3$ operator the above procedure gives
\eqn\grav{\eqalign{V_{s=3}=ce^{-3\phi}\partial{X^{a_1}}\partial{X^{a_2}}
\psi^{a_3}{e^{i{\vec{p}}{\vec{X}}}}H_{a_1a_2a_3}(p)
\rightarrow\oint{dz}V_1\cr
=
H_{a_1a_2a_3}(p)
{\oint}e^{\phi}\partial{X^{a_1}}\partial{X^{a_2}}
\psi^{a_3}{e^{i{\vec{p}}{\vec{X}}}}\cr
\lbrack{Q_1},V_1\rbrack=\partial{U}=
H_{a_1a_2a_3}(p)\partial(c
e^{\phi}\partial{X^{a_1}}\partial{X^{a_2}}
\psi^{a_3}{e^{i{\vec{p}}{\vec{X}}}})\cr
\lbrack{Q_2},V_1\rbrack=W_1={1\over2}H_{a_1a_2a_3}(p)e^{2\phi-\chi}
\lbrace({-}({\vec{\psi}}\partial{\vec{X}})
+i({\vec{p}}{\vec{\psi}})P^{(1)}_{\phi-\chi}+i
({\vec{p}}\partial{\vec{\psi}}))
\partial{X^{a_1}}\partial{X^{a_2}}
\psi^{a_3}{e^{i{\vec{p}}{\vec{X}}}}\cr
+\partial{X^{a_1}}(\partial^2\psi^{a_2}+2\partial\psi^{a_2}P^{(1)}_{\phi-\chi})
\psi^{a_3}-
\partial{X^{a_1}}\partial{X^{a_2}}(\partial^2{X^{a_3}}+
\partial{X^{a_3}}P^{(1)}_{\phi-\chi})\rbrace{e^{i{\vec{p}}{\vec{X}}}}
\cr
\lbrack{Q_3},V_1\rbrack=W_2=-{1\over4}H_{a_1a_2a_3}(p)e^{3\phi-2\chi}
P^{(1)}_{2\phi-2\chi-\sigma}
\partial{X^{a_1}}\partial{X^{a_2}}
\psi^{a_3}{e^{i{\vec{p}}{\vec{X}}}}}}
where the conformal weight $n$ polynomials in the derivatives
of the ghost fields $\phi,\chi, \sigma$ are defined according
to ~{\selfb, \selfc}:

\eqn\lowen{P^{(n)}_{f(\phi,\chi,\sigma)}=e^{-f(\phi(z),\chi(z),\sigma(z))}
{{\partial^{n}}\over{\partial{z^n}}}e^{f(\phi(z),\chi(z),\sigma(z))}}
where $f$ is some linear function in $\phi,\chi,\sigma$.
For example, $P^{(1)}_{\phi-\chi}=\partial\phi-\partial\chi$, etc.
Note that the  product (43) is defined in the algebraic sense 
(not as an operator product).

Accordingly,
\eqn\grav{\eqalign{:K\partial^2{W_1}:
=4H_{a_1a_2a_3}(p)c\xi
\lbrace({-}({\vec{\psi}}\partial{\vec{X}})
+i({\vec{p}}{\vec{\psi}})P^{(1)}_{\phi-\chi}+i
({\vec{p}}\partial{\vec{\psi}}))
\partial{X^{a_1}}\partial{X^{a_2}}
\psi^{a_3}{e^{i{\vec{p}}{\vec{X}}}}\cr
+\partial{X^{a_1}}(\partial^2\psi^{a_2}+2\partial\psi^{a_2}P^{(1)}_{\phi-\chi})
\psi^{a_3}-
\partial{X^{a_1}}\partial{X^{a_2}}(\partial^2{X^{a_3}}+
\partial{X^{a_3}}P^{(1)}_{\phi-\chi})\rbrace{e^{i{\vec{p}}{\vec{X}}}}\cr
:K\partial^2W_2:=H_{a_1a_2a_3}(p){\lbrace}-\partial^2(
e^{\phi}\partial{X^{a_1}}\partial{X^{a_2}}
\psi^{a_3}{e^{i{\vec{p}}{\vec{X}}}})+
P^{(2)}_{2\phi-2\chi-\sigma}
e^{\phi}\partial{X^{a_1}}\partial{X^{a_2}}
\psi^{a_3}{e^{i{\vec{p}}{\vec{X}}}}\rbrace}}
and
\eqn\grav{\eqalign{
:\partial^{2n+1}K{K}\lbrace{Q_{brst}},U\rbrace:=
-24H_{a_1a_2a_3}(p)\partial{c}c\partial\xi\xi{e^{-\phi}}
\partial{X^{a_1}}\partial{X^{a_2}}
\psi^{a_3}{e^{i{\vec{p}}{\vec{X}}}}\cr
:\partial^{m}K{K}\lbrace{Q_{brst}},U\rbrace:=0 (m<2n+1)}}
and therefore, upon integrating out total derivatives,
 the  complete BRST-invariant expression
for the $s=3$ operator at picture 1 is
\eqn\grav{\eqalign{A_{s=3}(w)=
H_{a_1a_2a_3}(p)\oint{dz}(z-w)^2\lbrace
{1\over2}
P^{(2)}_{2\phi-2\chi-\sigma}
e^{\phi}\partial{X^{a_1}}\partial{X^{a_2}}
\psi^{a_3}
\cr+
2c\xi
\lbrack({-}({\vec{\psi}}\partial{\vec{X}})
+i({\vec{p}}{\vec{\psi}})P^{(1)}_{\phi-\chi}+i
({\vec{p}}\partial{\vec{\psi}}))
\partial{X^{a_1}}\partial{X^{a_2}}
\psi^{a_3}{e^{i{\vec{p}}{\vec{X}}}}\cr
+\partial{X^{a_1}}(\partial^2\psi^{a_2}+2\partial\psi^{a_2}P^{(1)}_{\phi-\chi})
\psi^{a_3}-
\partial{X^{a_1}}\partial{X^{a_2}}(\partial^2{X^{a_3}}+
\partial{X^{a_3}}P^{(1)}_{\phi-\chi})\rbrack
\cr
-12\partial{c}c\partial\xi\xi{e^{-\phi}}
\partial{X^{a_1}}\partial{X^{a_2}}
\psi^{a_3}\rbrace{e^{i{\vec{p}}{\vec{X}}}}}}

To abbreviate notations for our calculations of the correlation
functions in the following sections, it is convenient
to write the vertex operator $A_{s=3}$ (46) as a sum 

\eqn\lowen{A_{s=3}=A_0+A_1+A_2+A_3+A_4+A_5+A_6}
where
\eqn\lowen{A_0(w)={1\over2}H_{a_1a_2a_3}(p)\oint{dz}(z-w)^2
P^{(2)}_{2\phi-2\chi-\sigma}
e^{\phi}\partial{X^{a_1}}\partial{X^{a_2}}
\psi^{a_3}{e^{i{\vec{p}}{\vec{X}}}}(z)}
and
\eqn\lowen{A_6(w)=-12H_{a_1a_2a_3}(p)\oint{dz}(z-w)^2
\partial{c}c\partial\xi\xi{e^{-\phi}}
\partial{X^{a_1}}\partial{X^{a_2}}
\psi^{a_3}\rbrace{e^{i{\vec{p}}{\vec{X}}}}(z)}
have ghost factors proportional to
$e^\phi$ and $\partial{c}c\partial\xi\xi{e^{-\phi}}$ respectively
and the rest of the terms carry ghost factor proportional to
$c\xi$:
\eqn\grav{\eqalign{A_1(w)=-2H_{a_1a_2a_3}(p)\oint{dz}(z-w)^2
c\xi({\vec{\psi}}\partial{\vec{X}})
\partial{X^{a_1}}\partial{X^{a_2}}
\psi^{a_3}
{e^{i{\vec{p}}{\vec{X}}}}(z)\cr
A_2(w)=
2H_{a_1a_2a_3}(p)\oint{dz}(z-w)^2
c\xi(\partial^2\psi^{a_2}+2\partial\psi^{a_2}P^{(1)}_{\phi-\chi})
\psi^{a_3}
{e^{i{\vec{p}}{\vec{X}}}}(z)\cr
A_3(w)=
-2H_{a_1a_2a_3}(p)\oint{dz}(z-w)^2c\xi
\partial{X^{a_1}}\partial{X^{a_2}}(\partial^2{X^{a_3}}+
\partial{X^{a_3}}P^{(1)}_{\phi-\chi})
{e^{i{\vec{p}}{\vec{X}}}}(z)\cr
A_4(w)=2iH_{a_1a_2a_3}(p)\oint{dz}(z-w)^2
c\xi({\vec{p}}{\vec{\psi}})P^{(1)}_{\phi-\chi}
\partial{X^{a_1}}\partial{X^{a_2}}
\psi^{a_3}
{e^{i{\vec{p}}{\vec{X}}}}(z)\cr
A_5(w)=
2iH_{a_1a_2a_3}(p)\oint{dz}(z-w)^2
c\xi({\vec{p}}\partial{\vec{\psi}})
\partial{X^{a_1}}\partial{X^{a_2}}
\psi^{a_3}
{e^{i{\vec{p}}{\vec{X}}}}(z)
}}
We are now prepared to analyze the $4$-point $1-1-3-3$
amplitude (leading to the gauge-invariant quartic 
 interaction of spin 3 and spin 1 particles),
which will be computed in the next sections.

\centerline{\bf $1-1-3-3$ Quartic Potential - preliminaries}

The goal of next two sections is to compute
 the 4-point function
of two $s=3$ vertex operators with $2$-photons, describing the gauge-invariant
$1-1-3-3$ interactions in the low energy limit of string theory.
The photon vertex operators are the standard ones, and it is convenient
to take them unintegrated at superconformal pictures $-1$ and $-2$ :
\eqn\grav{\eqalign{
V_{s=1}^{(-1)}(p)=c{e^{-\phi}}\psi^{m}e^{i{\vec{p}}{\vec{X}}}A_m(p)
\cr
V_{s=1}^{(-2)}(p)=c{e^{-2\phi}}\partial{X^{m}}e^{i{\vec{p}}{\vec{X}}}A_m(p)
}}
To cancel the background charges, the operators  in the 4-point
3-3-1-1 amplitude must be chosen  to have
 total $b-c$ ghost number $+3$, $\phi$-ghost number $-2$ and
$\chi$-ghost number $+1$. 
Therefore, with the picture choice (33) for the photons
it is 
clear that both of the $s=3$ operators have to be 
taken at their positive picture $+1$ representation (32). It is
furthermore clear that the amplitude
$A(1-1-3-3)(p_1,...,p_4)$
 will only be contributed
by the terms:
\eqn\grav{\eqalign{
A(1-1-3-3)(p_1,...,p_4)=S(1-1-3-3)(p_1,...,p_4)+(p_3{\leftrightarrow}p_4)
\cr
S(1-1-3-3)\equiv
<V_{s=1}(p_1)V_{s=1}(p_2)V_{s=3}(p_3)V_{s=3}(p_4)>
\cr
=\sum_{j=1}^5<V_{s=1}(p_1)V_{s=1}(p_2)A_j(p_3)A_0(p_4)>+(p_3\leftrightarrow
p_4)}}
with$ A_0,A_j$ given in (32). Note that, with the picture choice
(33) for the $s=1$ operators, the $A_6$-part of
$V_{s=3}$ at positive picture does not contribute to the correlator
at all
due  to the ghost balance constraint.
The structure of the amplitude (34) is remarkably different
from the standard Veneziano form. Recall that the standard
Veneziano expression for $4$-point amplitude in string theory
arises as a result of 3 out of 4 operators taken 
unintegrated (multiplied by the c-ghosts) and one integrated
(with the $b-c$ ghost number 0), in order to ensure the $b-c$ ghost anomaly
cancellation ( this choice is related to fixing the
$SL(2,R)$ global symmetry with the ghost part of the correlator
producing the standard Koba-Nielsen's determinant).
The single integration then leads to the Veneziano structures
$\sim{{\Gamma\Gamma}\over{\Gamma}}$ in the open string case or
$\sim{{\Gamma\Gamma\Gamma}\over{\Gamma\Gamma\Gamma}}$ for closed strings
where $\Gamma$ are the gamma-functions of Mandelstam variables.
With the $s=3$ vertex operators the structure
of amplitudes is different,
as their ghost couplings (both $b-c$ and $\beta-\gamma$)
are nonstandard. 
For example, the $s=3$ operators at positive pictures exist
in the  integrated form only (unlike the standard operators
that can be taken integrated or unintegrated); at the same time,
their integrands contain terms with $b-c$ ghost numbers 1 and 2
(as opposed to the standard integrated vertices which integrands
have ghost number zero).
As it is clear from (32)-(34) the ghost number balance
of the $1-1-3-3$ four-point function requires both of
the $s=3$ operators to be taken integrated at positive pictures.
Therefore the 4-point amplitude involves the double worldsheet integration
and its form is quite different from Veneziano type.
In particular, it leads to nonlocalities appearing
in the quartic interactions involving the higher spins.
Our goal now is to analyze the $<VVA_jA_0>$-correlators contributing
to the 4-point amplitude (34) one by one.
The first step is to fix
the points $u_1,u_2,w_1,w_2$ in the
amplitude $<V_{s=1}(u_1)V_{s=1}(u_2)A_{s=3}(w_1)A_{s=3}(w_2)>$
by using
 the remnant gauge symmetry on the worldsheet.
Note that, while
$u_1,u_2$  are the actual points of the unintegrated $s=1$ vertices,
$w_1$ and $w_2$ are the points defining the  contours
for the integrated $s=3$ vertices at positive pictures
(corresponding to the  $w$-points  
in the expression (32) for the $A_{s=3}$-vertex).
In the standard N-point amplitude case (involving 3 unintegrated vertices
and $N-3$ integrated) the remnant gauge symmetry is well-known to be
given by $SL(2,R)$ subgroup of conformal symmetry, allowing to fix
the locations of the unintegrated operators at 
3 particular points (with the standard choice
$0,1$ and $\infty$). In the operator language, the $SL(2,R)$ symmetry
simply reflects the fact that, translating an unintegrated vertex
operator of the form $\sim{cV}(z_1)$ to some new point $z_2$
changes it by BRST-exact terms not contributing to correlation functions
(since all the $z$-derivatives of the unintegrated vertices are BRST-exact,
e.g. $\partial(cV)(z)=\lbrack{Q},V(z)\rbrack $ etc.
In our case, because of the nonstandard ghost structure of the
spin 3 operators, the situation is different and the actual
remnant gauge symmetry is bigger than $SL(2,R)$.
Namely, all $w$-derivatives of the $A_{s=3}(w)$ operators are BRST-exact,
so the $w$-points can be chosen arbitrarily.
So in case of the $4$-point amplitude (34) the remnant gauge symmetry on the 
worldsheet allows to fix 4 rather than 3 points, i.e. contains an
extra generator in addition to the standard $SL(2,R)$ part.
As it has been pointed out in ~{\selfc}, the extra gauge symmetries
on the worldsheet are closely related to the $global$ space-time 
$\alpha$-symmetries that are realized nonlinearly and stem
from hidden space-time dimensions in string theory. Just
like the higher spin vertices, the $\alpha$-symmetry generators
are essentially mixed with the ghosts, being the elements
of nontrivial ghost cohomologies $H_{-3}\sim{H_1},H_{-4}\sim{H_2}$
and $H_{-5}\sim{H_3}$ with each cohomology essentially contributing
an extra space-time dimension. In this context, as the higher spin
vertex operators and the $\alpha$-symmetries have similar ghost
cohomology structures, the appearance of extra gauge symmetries
on the worldsheet is not surprising.
Therefore using the $SL(2,R)$ symmetries plus the extra symmetry
it is convenient to set 
\eqn\grav{\eqalign{
z_1=0, z_2=\infty \cr
w_1=w_2=0}}

Such a choice may appear somewhat unusual; indeed, in the standard
case the unintegrated vertices
are set at three different points (e.g. such as
$0,1,\infty$), 
 since, if one formally fixes two operators
at  coincident (or infinitely close)
 points, one faces the normal ordering issue
(although the $SL(2,R)$ symmetry in principle
allows to fix the operators at 3 infinitely close points)
In case of the higher spin operators, however, 
fixing the $w$-points is merely related to the choice of their 
integration contours,thus
the gauge choice (35) is appropriate.

\centerline{\bf $3-3-1-1$ Amplitude: the calculation}
It is convenient to start with evaluating the ghost part
of the 4-point function, common
for all the terms in (34).
We get
\eqn\grav{\eqalign{F_{gh}(u,z_1,z_2)=
lim_{u\rightarrow{\infty}}<c{e^{-2\phi}}(0)c{e^{-\phi}}(u)
c{e^\chi}(z_1)P^{(2)}_{2\phi-2\chi-\sigma}e^\phi(z_2)>
\cr
={{6uz_1(z_1^2+z_2^2)}\over{(z_1-z_2)^2}}}}

The first contribution is given by
\eqn\grav{\eqalign{
S_1(1-1-3-3)=
<V_{s=1}(p_1;0)V_{s=1}(p_2;\infty)A_1(p_3;0)
A_0(p_4,0)>\cr
= 
A_m(p_1)A_n(p_2)H_{a_2a_3a_4}(p_3)H_{b_1b_2b_3}(p_4)
lim_{u\rightarrow\infty}\int_{0}^{1}{dz_2}\int_{0}^{z_2}{dz_1}z_1^2z_2^2
F_{gh}(u,z_1,z_2)
\cr\times
<\partial{X^m}e^{i{\vec{p_1}}{\vec{X}}}(0)
\psi^ne^{i{\vec{p_2}}{\vec{X}}}(u)
(\psi_{a_1}\partial{X^{a_1}})\psi^{a_4}\partial{X^{a_2}}
\partial{X^{a_3}}e^{i{\vec{p_3}}{\vec{X}}}
\partial{X^{b_1}}\partial{X^{b_2}}\psi^{b_3}e^{i{\vec{p_4}}{\vec{X}}}
>}}
The $\psi$-correlator gives
\eqn\grav{\eqalign{
lim_{u\rightarrow\infty}
<\psi^n(u)\psi^{a_1}\psi^{a_4}(z_1)\psi^{b_3}(z_2)>
={{\eta^{na_1}\eta^{a_4b_3}-\eta^{na_4}\eta^{a_1b_3}}\over{u(z_1-z_2)}}}}
so the $\psi$-correlator multiplied by $F_{gh}(u,z_1,z_2)$
gives ${{6(\eta^{na_1}\eta^{a_4b_3}-\eta^{na_4}\eta^{a_1b_3})z_1
(z_1+z_2)^2}\over{(z_1-z_2)^3}}$
with the $u$-factor cancelled.
Due to conformal invariance, it is clear that the remaining $X$-correlator
will contribute terms of the order of $u^0$ to the overall correlator, 
with all other terms vanishing on-shell - in other words, no pairings
of $\partial{X}$'s with $e^{{\vec{p_2}}{\vec{X}}}$ contribute to the overall
4-point amplitude. For this reason, the 
relevant contributions from  the $X$-correlator are reduced to
the three-point function
\eqn\grav{\eqalign{S_X=
<\partial{X^m}e^{{\vec{p_1}}{\vec{X}}}(0)
\partial{X^{a_1}}\partial{X^{a_2}}\partial{X^{a_3}}
e^{{\vec{p_3}}{\vec{X}}}(z_1)
\partial{X^{b_1}}\partial{X^{b_2}}\partial{X^{b_3}}
e^{{\vec{p_4}}{\vec{X}}}(z_2)>}}
This function is not difficult to evaluate.
To keep our expressions
as compact as possible for the subsequent integrations in $z_1,z_2$,
it is convenient to use the following notations for
computing the $X$-correlators.

Namely, each term contributing to the correlator (39) can be classified
in terms of numbers of pairings between $\partial{X}$'s with the
exponents and between each other.
That is,
let $M_1,M_2$ be  pairing numbers
between $\partial{X_m}(0)$ and $e^{{\vec{p_3}}{\vec{X}}}(z_1),
e^{{\vec{p_4}}{\vec{X}}}(z_2)$ respectively with the obvious constraint
$0\leq{M_1,M_2}=1$ (since there is only 
one $\partial{X}$ in the expression for the
photon. Next, let
$N_1,N_2$ be pairing numbers of $\partial{X}$'s in the $s=3$ operator
at $z_1$ with $e^{{\vec{p_1}}{\vec{X}}}(0)$ and $e^{{\vec{p_4}}{\vec{X}}}(z_2)$
with  $0{\leq}N_1,N_2{\leq}3$. Finally
$P_1,P_2$ satisfying $0{\leq}P_1,P_2{\leq}2$ shall stand for the pairings
between $\partial{X}$'s of the second $s=3$ vertex at $z_2$ with 
 $e^{{\vec{p_1}}{\vec{X}}}(0)$ and $e^{{\vec{p_3}}{\vec{X}}}(z_1)$
It is then straightforward to show that the correlator is
contributed by two types of terms. The first type includes the kinematic
factors sextic in momentum (accordingly, leading to six derivative interactions
in the low energy limit). These terms appear when all $\partial{X}$'s
in the correlator (39) (total number 6) are contracted with the exponents.
The second type ivolves the kinematic factors quartic in momentum,
appearing when 4 out of 6 $\partial{X}$'s are contracted with 
the exponents, while the remaining two are contracted with each other.
These terms lead to four derivative quartic interactions in space-time.
Given the Pauli-Fierz conditions (3) on the $s=3$ fields, there are no 
terms quadratic in momentum or momentum-independent.

Computing the $X$-correlator (39) and multiplying by the $\psi$-ghost
factor (38), we obtain
the six-derivative part of the correlator $S_1^{6-der}(1-1-3-3)$ (37),  given by

\eqn\grav{\eqalign{S_1^{6-der}(1-1-3-3)=
72A_m(p_1)A_n(p_2)H_{a_2a_3a_4}(p_3)H_{b_1b_2b_3}(p_4)
\cr\times
({{\eta^{na_1}\eta^{a_4b_3}-\eta^{na_4}\eta^{a_1b_3}}})
\sum_{M_1=0}^1
\sum_{N_1=0}^3\sum_{P_1=0}^2
{{(-1)^{P_1}}\over{N_1!(3-N_1)!P_1!(2-P_1)!}}
\cr\times
\prod_{\alpha=1}^{n_1}\prod_{\beta=N_1+1}^3
\prod_{\gamma=1}^{P_1}\prod_{\lambda=P_1+1}^3(ip_1)^{a_\alpha}(ip_4)^{a_\beta}
(ip_1)^{b_\gamma}(ip_3)^{b_\lambda}(ip_3^m)^{M_1}(ip_4^m)^{1-M_1}
\cr\times
\int_{0}^{1}{dz_2}\int_{0}^{z_2}{dz_1}z_1^2z_2^2(z_1^2+z_2^2)
z_1^{1+{{\vec{p}}_1{\vec{p}}_3-M_1-N_1}}
z_2^{{{\vec{p}}_1{\vec{p}}_4-1+M_1-P_1}}(z_1-z_2)^{{{\vec{p}}_3{\vec{p}}_4-8+N_1+P_1}}
}}

Few comments should be made to explain our notations here and below.
Firstly, regarding the products appearing in (40):
for example, $\prod_{\alpha=1}^{N_1}(ip_1)^{a_\alpha}$ stands for the
usual product
$(ip_1^{a_1})...(ip_1^{a_{N_1}})$ for $1\leq{N_1}\leq{3}$, but
is set to $1$ if $N_1=0$.
Similarly,
$\prod_{\beta=N_1+1}^3(ip_4^{a_\beta})$ stands for the product
$ip_4^{a_{N_1+1}}...ip_4^{a_3}$ if $N_1=0,1,2$ but is set to 1
if $N_1=3$. Similarly for all other products of that type.
The product 
$(ip_3^m)^{M_1}(ip_4^m)^{1-M_1}$ obviously stands for $ip_4^m$  for
$M_1=0$  and $ip_3^m$ for $M_1=1$ and similarly for all other products
of that type.
The next step is to perform the integration
of (37) in $z_1$ and $z_2$.
This can be done by
using 
\eqn\grav{\eqalign{
\int_{0}^{1}{dz_2}\int_{0}^{z_2}{dz_1}z_1^a{z_2^b}(z_1-z_2)^c
={{\Gamma(a+1)\Gamma(c+1)}\over{(a+b+c+2)\Gamma(a+c+2)}}}}
Integrating (37) then gives the following answer for
 $S_1^{6-der}(1-1-3-3)$:
\eqn\grav{\eqalign{
S_1^{6-der}(1-1-3-3)=
72A_m(p_1)A_n(p_2)H_{a_2a_3a_4}(p_3)H_{b_1b_2b_3}(p_4)
\cr\times
({{\eta^{na_1}\eta^{a_4b_3}-\eta^{na_4}\eta^{a_1b_3}}})
\sum_{M_1=0}^1
\sum_{N_1=0}^3\sum_{P_1=0}^2
{{(-1)^{P_1}}\over{N_1!(3-N_1)!P_1!(2-P_1)!}}
\cr\times
\prod_{\alpha=1}^{n_1}\prod_{\beta=N_1+1}^3
\prod_{\gamma=1}^{P_1}\prod_{\lambda=P_1+1}^3(ip_1)^{a_\alpha}(ip_4)^{a_\beta}
(ip_1)^{b_\gamma}(ip_3)^{b_\lambda}(ip_3^m)^{M_1}(ip_4^m)^{1-M_1}
({\vec{p_3}}{\vec{p_4}}-{\vec{p_1}}{\vec{p_2}})^{-1}
\cr\times
\Gamma({\vec{p_3}}{\vec{p_4}}+N_1+P_1-7)
\lbrack
{{\Gamma({\vec{p_1}}{\vec{p_3}}-M_1-N_1+6)}\over
{\Gamma(-{\vec{p_2}}{\vec{p_3}}-M_1+P_1-1)}}
+
{{\Gamma({\vec{p_1}}{\vec{p_3}}-M_1-N_1+4)}\over
{\Gamma(-{\vec{p_2}}{\vec{p_3}}-M_1+P_1-3)}}
\rbrack}}
We find that the expression (42)
contains the factor
$G(p_1,p_2,p_3,p_4)=({\vec{p_3}}{\vec{p_4}}-{\vec{p_1}}{\vec{p_2}})^{-1}$
(this factor will actually appear in all the terms in the 4-point
amplitude (34)).
If we are on-shell, the denominator in this expression is zero and
the correlator (42) diverges. It must be stressed, however,
that terms in the low-energy effective action, appearing in the field theory
limit of string theory, are determined by appropriate
terms in
 conformal beta-functions on the worldsheet, rather than by the 
on-shell correlators. The conformal beta-function, in turn, is
determined by the structure constants that are essentially taken off-shell
(the on-shell limit then corresponds to the constraint $\beta=0$).
For example, if $\Phi$ is a scalar massless space-time field,
to obtain linear term in its $\beta$-function proportional to
$\sim{\Delta}\Phi=-p^2\Phi$ (corresponding to the free field part
of its low energy effective action), one has to take 
the dilaton's vertex operator initially off-shell (so
that $p^2\neq{0}$) and perform the internal normal ordering in this
vertex operator leading to the flow $\sim{p^2\Phi{log}\Lambda}$
where $\Lambda$ is the worldsheet cutoff.
Similarly, the denominator of $G(p_1,...,p_4)$ is nonzero 
 in the off-shell limit relevant to the $\beta$-function computations,
so the corresponding quartic terms in the low-energy
effective action include the factor
\eqn\lowen{G(p_1,p_2,p_3,p_4)=(p_1^2+p_2^2-p_3^2-p_4^2)^{-1}}
where we used $(p_1+p_2)^2=(p_3+p_4)^2$.
This is the factor reflecting the nonlocality of the quartic
couplings of the higher spin fields in the position space.
We find that, from the string theory point of view,
this nonlocality is the consequence of the specific ghost structure
of the higher spin vertex operators, as we already noted above.
The calculation
of the $4-derivative$ part of the correlator (37) (quartic in momentum)
is similar. The result is given by
\eqn\grav{\eqalign{
S_1^{4-der}(1-1-3-3)=D_1+D_2+D_3}}
where
\eqn\grav{\eqalign{
D_1=
-72A_m(p_1)A_n(p_2)H_{a_1a_2a_4}(p_3)H_{b_1b_2b_3}(p_4)
({{\eta^{na_3}\eta^{a_4b_3}-\eta^{na_4}\eta^{a_3b_3}}})
\eta^{ma_3}
\cr\times
\sum_{N_1=0}^2\sum_{P_1=0}^2
{{(-1)^{P_1}}\over{N_1!(2-N_1)!P_1!(2-P_1)!}}
\prod_{\alpha=1}^{n_1}\prod_{\beta=N_1+1}^3
\prod_{\gamma=1}^{P_1}\prod_{\lambda=P_1+1}^3(ip_1)^{a_\alpha}(ip_4)^{a_\beta}
(ip_1)^{b_\gamma}(ip_3)^{b_\lambda}
\cr\times
G(p_1,p_2,p_3,p_4)
\Gamma({\vec{p_3}}{\vec{p_4}}+N_1+P_1-6)
\lbrack
{{\Gamma({\vec{p_1}}{\vec{p_3}}-N_1+4)}\over
{\Gamma(-{\vec{p_2}}{\vec{p_3}}+P_1-2)}}
+
{{\Gamma({\vec{p_1}}{\vec{p_3}}-N_1+2)}\over
{\Gamma(-{\vec{p_2}}{\vec{p_3}}+P_1-4)}}
\rbrack}}

\eqn\grav{\eqalign{
D_2=
-72A_m(p_1)A_n(p_2)H_{a_1a_2a_4}(p_3)H_{b_1b_2b_3}(p_4)
({{\eta^{na_3}\eta^{a_4b_3}-\eta^{na_4}\eta^{a_3b_3}}})
\eta^{mb_2}
\cr
\sum_{N_1=0}^3\sum_{P_1=0}^1
{{(-1)^{P_1}}\over{N_1!(3-N_1)!}}
\prod_{\alpha=1}^{N_1}\prod_{\beta=N_1+1}^3
(ip_1)^{a_\alpha}(ip_4)^{a_\beta}
(ip_1^{b_1})^{P_1}(ip_3^{b_1})^{1-P_1}
\cr\times
G(p_1,p_2,p_3,p_4)
\Gamma({\vec{p_3}}{\vec{p_4}}+N_1+P_1-6)
\lbrack
{{\Gamma({\vec{p_1}}{\vec{p_3}}-N_1+6)}\over
{\Gamma(-{\vec{p_2}}{\vec{p_3}}+P_1)}}
+
{{\Gamma({\vec{p_1}}{\vec{p_3}}-N_1+4)}\over
{\Gamma(-{\vec{p_2}}{\vec{p_3}}+P_1-2)}}
\rbrack}}
\eqn\grav{\eqalign{
D_3=
-72A_m(p_1)A_n(p_2)H_{a_1a_2a_4}(p_3)H_{b_1b_2b_3}(p_4)
\cr\times
({{\eta^{na_3}\eta^{a_4b_3}-\eta^{na_4}\eta^{a_3b_3}}})
\eta^{mb_2}
\sum_{M_1=0}^1\sum_{N_1=0}^2\sum_{P_1=0}^1
{{(-1)^{P_1}}\over{N_1!(2-N_1)!}}
\cr\times
\prod_{\alpha=1}^{N_1}\prod_{\beta=N_1+1}^3
(ip_1)^{a_\alpha}(ip_4)^{a_\beta}
(ip_3^m)^{M_1}(ip_4^m)^{1-M_1}
(ip_1^{b_1})^{P_1}(ip_3^{b_1})^{1-P_1}
\cr\times
G(p_1,p_2,p_3,p_4)
\Gamma({\vec{p_3}}{\vec{p_4}}+N_1+P_1-7)
\lbrack
{{\Gamma({\vec{p_1}}{\vec{p_3}}-M_1-N_1+6)}\over
{\Gamma(-{\vec{p_2}}{\vec{p_3}}-M_1+P_1-1)}}
\cr
+
{{\Gamma({\vec{p_1}}{\vec{p_3}}-M_1-N_1+4)}\over
{\Gamma(-{\vec{p_2}}{\vec{p_3}}+P_1-M_1-3)}}
\rbrack}}
This concludes the computation of
$S_1(1-1-3-3)$ contribution to the quartic interaction of
$1-1-3-3$.
The next contribution, $S_2({1-1-3-3})$,
is given by
\eqn\grav{\eqalign{
S_2(1-1-3-3)
=
<V_{s=1}(p_1;0)V_{s=1}(p_2;\infty)A_2(p_3;0)
A_0(p_4,0)>\cr
= 
A_m(p_1)A_n(p_2)H_{a_1a_2a_3}(p_3)H_{b_1b_2b_3}(p_4)
lim_{u\rightarrow\infty}\int_{0}^{1}{dz_2}\int_{0}^{z_2}{dz_1}z_1^2z_2^2
\cr\times
<ce^{-2\phi}\partial{X^m}e^{i{\vec{p_1}}{\vec{X}}}(0)
ce^{-\phi}\psi^ne^{i{\vec{p_2}}{\vec{X}}}(u)ce^\chi
\partial{X^{a_1}}\psi^{a_2}(\partial^2\psi^{a_3}
\cr
+2\partial\psi^{a_3}
P^{(1)}_{\phi-\chi})e^{i{\vec{p_3}}{\vec{X}}}(z_3)
P^{(2)}_{2\phi-2\chi-\sigma}e^{\phi}
\partial{X^{b_1}}\partial{X^{b_2}}\psi^{b_3}e^{i{\vec{p_4}}{\vec{X}}}(z_2)
>}}
The $<\psi\times$ghost$>$  factor of this contribution is:
\eqn\grav{\eqalign{lim_{u\rightarrow\infty}<ce^{-2\phi}(0)ce^{-\phi}\psi^n(u)
ce^\chi\psi^{a_2}(\partial^2\psi^{a_3}
+2\partial\psi^{a_3}
P^{(1)}_{\phi-\chi})(z_1){e^\phi}P^{(2)}_{2\phi-2\chi-\sigma}\psi^{b_3}(z_2)>
\cr
=
{{24\eta^{na_2}\eta^{a_3b_3}z_1z_2^3(z_1^2+z_2^2)}\over{(z_1-z_2)^3}}+
O(u^{-1})
}}

Computing the $<X>$ part using the same conventions as above
and  integrating the overall correlator over $z_1$ and $z_2$
we obtain

\eqn\grav{\eqalign{
S_2(1-1-3-3)
\cr
=
48A_m(p_1)A_n(p_2)H_{a_1a_2a_3}(p_3)H_{b_1b_2b_3}(p_4)
\eta^{na_2}\eta^{a_3b_3}
\sum_{M_1=0}^1\sum_{N_1=0}^1\sum_{P_1=0}^2
{{(-1)^{P_1}}\over{P_1!(2-P_1)!}}
\cr\times
\prod_{\alpha=1}^{P_1}\prod_{\beta=P_1+1}^2
(ip_1^{b_\alpha})(ip_3^{b_\beta})
G(p_1,p_2,p_3,p_4)(ip_3^m)^{M_1}(ip_4^m)^{1-M_1}
(ip_1^{a_1})^{N_1}(ip_4^{a_1})^{1-N_1}
\cr\times
\Gamma({\vec{p_3}}{\vec{p_4}}+N_1+P_1-7)
\lbrack
{{\Gamma({\vec{p_1}}{\vec{p_3}}-M_1-N_1+5)}\over
{\Gamma(-{\vec{p_2}}{\vec{p_3}}+P_1-M_1-2)}}
+
{{\Gamma({\vec{p_1}}{\vec{p_3}}-M_1-N_1+3)}\over
{\Gamma(-{\vec{p_2}}{\vec{p_3}}+P_1-M_1-4)}}
\rbrack}}

This contribution is quartic in momentum.
The next contribtion is given by

\eqn\grav{\eqalign{
S_3(1-1-3-3)=
<V_{s=1}(p_1;0)V_{s=1}(p_2;\infty)A_2(p_3;0)
A_0(p_4,0)>\cr
= 
A_m(p_1)A_n(p_2)H_{a_1a_2a_3}(p_3)H_{b_1b_2b_3}(p_4)
lim_{u\rightarrow\infty}\int_{0}^{1}{dz_2}\int_{0}^{z_2}{dz_1}z_1^2z_2^2
\cr\times
<ce^{-2\phi}\partial{X^m}e^{i{\vec{p_1}}{\vec{X}}}(0)
ce^{-\phi}\psi^ne^{i{\vec{p_2}}{\vec{X}}}(u)
c{e^\chi}
\partial{X^{a_1}}\partial{X^{a_2}}(\partial^2{X^{a_3}}+
\partial{X^{a_3}}P^{(1)}_{\phi-\chi})
{e^{i{\vec{p_3}}{\vec{X}}}}(z_1)
\cr
P^{(2)}_{2\phi-2\chi-\sigma}e^{\phi}
\partial{X^{b_1}}\partial{X^{b_2}}\psi^{b_3}e^{i{\vec{p_4}}{\vec{X}}}(z_2)
>}}
The computation gives:
\eqn\grav{\eqalign{
S_3(1-1-3-3)=S_3^{(1)}+S_3^{(2)}+S_3^{(3)}}}
where $S_3^{(1)}$ and $S_3^{(2)}$ are the contributions
quartic in momentum while
$S_3^{(3)}=S_3^{(3)4-der}+S_3^{(3)6-der}$ contains both 4 and 6 derivative terms.
These contributions are given by, accordingly:
\eqn\grav{\eqalign{
S_3^{(1)}=
24A_m(p_1)A_n(p_2)H_{a_1a_2a_3}(p_3)H_{b_1b_2b_3}(p_4)
\cr\times
\eta^{nb_3}\eta^{ma_3}
\sum_{N_1=0}^2\sum_{P_1=0}^2
{{(-1)^{P_1}}\over{P_1!(2-P_1)!N_1!(2-N_1)!}}
\cr\times
\prod_{\alpha=1}^{N_1}\prod_{\beta=N_1+1}^2
\prod_{\gamma=1}^{P_1}\prod_{\lambda={P_1+1}}^2
(ip_1^{a_\alpha})(ip_4^{a_\beta})(ip_1^{b_\gamma})(ip_3^{b_\lambda})
G(p_1,p_2,p_3,p_4)
\cr\times
\Gamma({\vec{p_3}}{\vec{p_4}}+N_1+P_1-6)
\lbrack
{{\Gamma({\vec{p_1}}{\vec{p_3}}-N_1+4)}\over
{\Gamma(-{\vec{p_2}}{\vec{p_3}}+P_1-2)}}
+
{{\Gamma({\vec{p_1}}{\vec{p_3}}-N_1+2)}\over
{\Gamma(-{\vec{p_2}}{\vec{p_3}}+P_1-4)}}
\rbrack}}

\eqn\grav{\eqalign{
S_3^{(2)}=
24A_m(p_1)A_n(p_2)H_{a_1a_2a_3}(p_3)H_{b_1b_2b_3}(p_4)
\eta^{nb_3}\eta^{ma_3}\sum_{M_1=0}^1
\sum_{N_1=0}^2\sum_{P_1=0}^1
{{(-1)^{P_1}}\over{N_1!(2-N_1)!}}
\cr\times
\prod_{\alpha=1}^{N_1}\prod_{\beta=N_1+1}^2
\prod_{\gamma=1}^{P_1}\prod_{\lambda={P_1+1}}^2
(ip_1^{a_\alpha})(ip_4^{a_\beta})(ip_3^m)^{M_1}(ip_4^m)^{1-M_1}
(ip_1^{b_1})^{P_1}(ip_3^{b_1})^{1-P_1}
G(p_1,p_2,p_3,p_4)
\cr\times
\Gamma({\vec{p_3}}{\vec{p_4}}+N_1+P_1-7)
\lbrack
{{\Gamma({\vec{p_1}}{\vec{p_3}}-M_1-N_1+6)}\over
{\Gamma(-{\vec{p_2}}{\vec{p_3}}-M_1+P_1-1)}}
+
{{\Gamma({\vec{p_1}}{\vec{p_3}}-M_1-N_1+2)}\over
{\Gamma(-{\vec{p_2}}{\vec{p_3}}-M_1+P_1-3)}}
\cr
+{{2\Gamma({\vec{p_1}}{\vec{p_3}}-M_1-N_1+5)}\over
{\Gamma(-{\vec{p_2}}{\vec{p_3}}-M_1+P_1-2)}}
+{{2\Gamma({\vec{p_1}}{\vec{p_3}}-M_1-N_1+3)}\over
{\Gamma(-{\vec{p_2}}{\vec{p_3}}-M_1+P_1-4)}}
\rbrack}}

The 6-derivative part of $S_3^{(3)}$ is given by
\eqn\grav{\eqalign{
S_3^{(3)6-der}=
24A_m(p_1)A_n(p_2)H_{a_1a_2a_3}(p_3)H_{b_1b_2b_3}(p_4)
\cr\times
\eta^{nb_3}\sum_{M_1=0}^1
\sum_{N_1=0}^2\sum_{P_1=0}^2\lbrace
{{(-1)^{P_1}}\over{P_1!(2-P_1)!N_1!(2-N_1)!}}
\cr\times
\prod_{\alpha=1}^{N_1}\prod_{\beta=N_1+1}^2
\prod_{\gamma=1}^{P_1}\prod_{\lambda={P_1+1}}^2
(ip_1^{a_\alpha})(ip_4^{a_\beta})(ip_1^{b_\gamma})(ip_3^{b_\lambda})
\cr\times
(ip_3^m)^{M_1}(ip_4^m)^{1-M_1}
G(p_1,p_2,p_3,p_4){\lbrace}
\cr\times
(ip_1)^{a_3}\Gamma({\vec{p_3}}{\vec{p_4}}+N_1+P_1-6)
\lbrack
{{\Gamma({\vec{p_1}}{\vec{p_3}}-M_1-N_1+4)}\over
{\Gamma(-{\vec{p_2}}{\vec{p_3}}-M_1+P_1-2)}}
\cr
+
{{\Gamma({\vec{p_1}}{\vec{p_3}}-M_1-N_1+2)}\over
{\Gamma(-{\vec{p_2}}{\vec{p_3}}-M_1+P_1-4)}}
\rbrack
\cr+(2ip_4^{a_3})
\Gamma({\vec{p_3}}{\vec{p_4}}+N_1+P_1-7)
\lbrack
{{\Gamma({\vec{p_1}}{\vec{p_3}}-M_1-N_1+5)}\over
{\Gamma(-{\vec{p_2}}{\vec{p_3}}-M_1+P_1-2)}}
\cr
+
{{\Gamma({\vec{p_1}}{\vec{p_3}}-M_1-N_1+3)}\over
{\Gamma(-{\vec{p_2}}{\vec{p_3}}-M_1+P_1-4)}}
\rbrack\rbrace}}
The 4-derivative part
of $S_3^{(3)}$
is given by

\eqn\grav{\eqalign{S_3^{(3)4-der}=
-24A_m(p_1)A_n(p_2)H_{a_1a_2a_3}(p_3)H_{b_1b_2b_3}(p_4)
\eta^{nb_3}\eta^{ma_3}
\sum_{N_1=0}^1\sum_{P_1=0}^2
{{(-1)^{P_1}}\over{P_1!(2-P_1)!}}
\cr\times
\prod_{\alpha=1}^{P_1}\prod_{\beta=P_1+1}^2
(ip_1^{b_\alpha})(ip_3^{b_\beta})(ip_1^{a_1})^{N_1}(ip_4^{a_1})^{1-N_1}
G(p_1,p_2,p_3,p_4)
\cr\times
\lbrace
ip_1^{a_3}\Gamma({\vec{p_3}}{\vec{p_4}}+N_1+P_1-5)
\lbrack
{{\Gamma({\vec{p_1}}{\vec{p_3}}-N_1+2)}\over
{\Gamma(-{\vec{p_2}}{\vec{p_3}}+P_1-3)}}
+
{{\Gamma({\vec{p_1}}{\vec{p_3}}-N_1)}\over
{\Gamma(-{\vec{p_2}}{\vec{p_3}}+P_1-5)}}
\rbrack
\cr
+
ip_4^{a_3}\Gamma({\vec{p_3}}{\vec{p_4}}+N_1+P_1-6)
\lbrack
{{\Gamma({\vec{p_1}}{\vec{p_3}}-N_1+3)}\over
{\Gamma(-{\vec{p_2}}{\vec{p_3}}+P_1-3)}}
+
{{\Gamma({\vec{p_1}}{\vec{p_3}}-N_1+1)}\over
{\Gamma(-{\vec{p_2}}{\vec{p_3}}+P_1-5)}}
\rbrack\rbrace
\cr
-24A_m(p_1)A_n(p_2)H_{a_1a_2a_3}(p_3)H_{b_1b_2b_3}(p_4)
\eta^{nb_3}\eta^{mb_2}
\sum_{N_1=0}^2\sum_{P_1=0}^1\lbrace
{{(-1)^{P_1}}\over{P_1!(2-P_1)!}}
\cr\times
\prod_{\alpha=1}^{N_1}\prod_{\beta=N_1+1}^2
(ip_1^{a_\alpha})(ip_4^{a_\beta})(ip_1^{b_1})^{P_1}(ip_3^{b_1})^{1-P_1}
G(p_1,p_2,p_3,p_4)
\cr\times
\lbrace
ip_1^{a_3}\Gamma({\vec{p_3}}{\vec{p_4}}+N_1+P_1-5)
\lbrack
{{\Gamma({\vec{p_1}}{\vec{p_3}}-N_1+4)}\over
{\Gamma(-{\vec{p_2}}{\vec{p_3}}+P_1-1)}}
+
{{\Gamma({\vec{p_1}}{\vec{p_3}}-N_1+2)}\over
{\Gamma(-{\vec{p_2}}{\vec{p_3}}+P_1-3)}}
\cr
+ip_4^{a_3}
\Gamma({\vec{p_3}}{\vec{p_4}}+N_1+P_1-6)
\lbrack
{{\Gamma({\vec{p_1}}{\vec{p_3}}-N_1+5)}\over
{\Gamma(-{\vec{p_2}}{\vec{p_3}}+P_1-1)}}
+
{{\Gamma({\vec{p_1}}{\vec{p_3}}-N_1+3)}\over
{\Gamma(-{\vec{p_2}}{\vec{p_3}}+P_1-3)}}
\rbrack
\rbrace
\cr
-24A_m(p_1)A_n(p_2)H_{a_1a_2a_3}(p_3)H_{b_1b_2b_3}(p_4)
\eta^{nb_3}\eta^{a_2b_2}
\sum_{M_1=0}^1\sum_{N_1=0}^1\sum_{P_1=0}^1
\cr\times
(ip_3^m)^{M_1}(ip_4^m)^{1-M_1}(ip_1^{a_1})^{N_1}(ip_4^{a_1})^{1-N_1}
(ip_1^{b_1})^{P_1}(ip_3^{b_1})^{1-P_1}
G(p_1,p_2,p_3,p_4)
\cr\times
\lbrace
ip_1^{a_3}\Gamma({\vec{p_3}}{\vec{p_4}}+N_1+P_1-6)
\lbrack
{{\Gamma({\vec{p_1}}{\vec{p_3}}-N_1-M_1+4)}\over
{\Gamma(-{\vec{p_2}}{\vec{p_3}}+P_1-M_1-2)}}
+
{{\Gamma({\vec{p_1}}{\vec{p_3}}-M_1-N_1+2)}\over
{\Gamma(-{\vec{p_2}}{\vec{p_3}}+P_1-M_1-4)}}
\cr
+ip_4^{a_3}
\Gamma({\vec{p_3}}{\vec{p_4}}+N_1+P_1-7)
\lbrack
{{\Gamma({\vec{p_1}}{\vec{p_3}}-N_1-M_1+5)}\over
{\Gamma(-{\vec{p_2}}{\vec{p_3}}+P_1-M_1-2)}}
+
{{\Gamma({\vec{p_1}}{\vec{p_3}}-N_1+3)}\over
{\Gamma(-{\vec{p_2}}{\vec{p_3}}+P_1-M_1-4)}}
\rbrack
\rbrace
}}
This concludes the computation of $S_3^{(3)}$ and
of $S_3(1-1-3-3)$.
The final contribution to the amplitude, $S_4(1-1-3-3)$,
is given by
\eqn\grav{\eqalign{S_4(1-1-3-3)=
<V_{s=1}(p_1;0)V_{s=1}(p_2;\infty)(A_4+A_5)(p_3;0)
A_0(p_4,0)>
\cr
=
iA_m(p_1)A_n(p_2)H_{a_1a_2a_3}(p_3)H_{b_1b_2b_3}(p_4)
lim_{u\rightarrow\infty}\int_{0}^{1}{dz_2}\int_{0}^{z_2}{dz_1}z_1^2z_2^2
\cr\times
<ce^{-2\phi}\partial{X^m}e^{i{\vec{p_1}}{\vec{X}}}(0)
ce^{-\phi}(\psi^ne^{i{\vec{p_2}}{\vec{X}}}(u)
c\xi({\vec{p}}{\vec{\psi}})P^{(1)}_{\phi-\chi}
+
{\vec{p}}\partial{\vec{\psi}})
\partial{X^{a_1}}\partial{X^{a_2}}
\psi^{a_3}
{e^{i{\vec{p_3}}{\vec{X}}}}(z_1)
\cr
P^{(2)}_{2\phi-2\chi-\sigma}e^{\phi}
\partial{X^{b_1}}\partial{X^{b_2}}\psi^{b_3}e^{i{\vec{p_4}}{\vec{X}}}(z_2)
>
}}
As previously, it is convenient
to split this contribution into 6 and 4 derivative parts:
\eqn\lowen{
S_4(1-1-3-3)=S_4^{6-der}+S_4^{4-der}
}
The 6-derivative part is computed to give
\eqn\grav{\eqalign{
S_4^{6-der}
=
-24iA_m(p_1)A_n(p_2)H_{a_1a_2a_3}(p_3)H_{b_1b_2b_3}(p_4)
\cr\times
\sum_{M_1=0}^1\sum_{N_1=0}^2\sum_{P_1=0}^2
{{(-1)^{P_1}}\over{N_1!(2-N_1)!P_1!(2-P_1)!}}
\cr\times
\prod_{\alpha=1}^{N_1}\prod_{\beta=N_1+1}^2
\prod_{\gamma=1}^{P_1}\prod_{\lambda=P_1+1}^2
(ip_1^{a_\alpha})(ip_4^{a_\beta})(ip_1^{b_\gamma})(ip_3^{b_\lambda})
(ip_3^m)^{M_1}(ip_3^m)^{1-M_1}
G(p_1,p_2,p_3,p_4)
\cr\times
\lbrace
2(\eta^{na_3}p_3^{b_3}+\eta^{a_3b_3}p_3^n)
\Gamma({\vec{p_3}}{\vec{p_4}}+N_1+P_1-7)
\lbrack
{{\Gamma({\vec{p_1}}{\vec{p_3}}-M_1-N_1+6)}\over
{\Gamma(-{\vec{p_2}}{\vec{p_3}}-M_1+P_1-1)}}
\cr
+
{{\Gamma({\vec{p_1}}{\vec{p_3}}-M_1-N_1+4)}\over
{\Gamma(-{\vec{p_2}}{\vec{p_3}}-M_1+P_1-3)}}
\rbrack
\cr
+
(2\eta^{na_3}p_3^{b_3}-\eta^{a_3b_3}p_3^n)
\Gamma({\vec{p_3}}{\vec{p_4}}+N_1+P_1-6)
\lbrack
{{\Gamma({\vec{p_1}}{\vec{p_3}}-M_1-N_1+5)}\over
{\Gamma(-{\vec{p_2}}{\vec{p_3}}-M_1+P_1-1)}}
\cr
+
{{\Gamma({\vec{p_1}}{\vec{p_3}}-M_1-N_1+3)}\over
{\Gamma(-{\vec{p_2}}{\vec{p_3}}-M_1+P_1-3)}}
\rbrack\rbrace
}}
Finally, the 4-derivative part of $S_4^{4-der}$
contributes
\eqn\lowen{S_4^{4-der}=S_4^{(1)4-der}+S_4^{(2)4-der}+S_4^{(3)4-der}}
where
\eqn\grav{\eqalign{
S_4^{(1)4-der}
=
24iA_m(p_1)A_n(p_2)H_{a_1a_2a_3}(p_3)H_{b_1b_2b_3}(p_4)
\sum_{N_1=0}^1\sum_{P_1=0}^2
{{(-1)^{P_1}}\over{P_1!(2-P_1)!}}
\cr\times
\prod_{\alpha=1}^{P_1}\prod_{\beta=P_1+1}^2
(ip_1^{b_\alpha})(ip_3^{b_\beta})(ip_1^{a_1})^{N_1}(ip_4^{a_1})^{1-N_1}
G(p_1,p_2,p_3,p_4)
\cr\times
\lbrace
2\eta^{a_2m}(\eta^{na_3}p_3^{b_3}+\eta^{a_3b_3}p_3^n)
\Gamma({\vec{p_3}}{\vec{p_4}}+N_1+P_1-5)
\lbrack
{{\Gamma({\vec{p_1}}{\vec{p_3}}-N_1+3)}\over
{\Gamma(-{\vec{p_2}}{\vec{p_3}}+P_1-2)}}
\cr
+
{{\Gamma({\vec{p_1}}{\vec{p_3}}-N_1+1)}\over
{\Gamma(-{\vec{p_2}}{\vec{p_3}}+P_1-4)}}
\rbrack
\cr
+
\eta^{a_2m}(2\eta^{na_3}p_3^{b_3}-\eta^{a_3b_3}p_3^n)
\Gamma({\vec{p_3}}{\vec{p_4}}+N_1+P_1-6)
\lbrack
{{\Gamma({\vec{p_1}}{\vec{p_3}}-N_1+4)}\over
{\Gamma(-{\vec{p_2}}{\vec{p_3}}+P_1-2)}}
\cr
+
{{\Gamma({\vec{p_1}}{\vec{p_3}}-N_1+2)}\over
{\Gamma(-{\vec{p_2}}{\vec{p_3}}+P_1-4)}}
\rbrack\rbrace}}
\eqn\grav{\eqalign{
S_4^{(2)4-der}
=
24iA_m(p_1)A_n(p_2)H_{a_1a_2a_3}(p_3)H_{b_1b_2b_3}(p_4)
\sum_{N_1=0}^2\sum_{P_1=0}^1
{{(-1)^{P_1}}\over{N_1!(2-N_1)!}}
\cr\times
\prod_{\alpha=1}^{N_1}\prod_{\beta=N_1+1}^2
(ip_1^{a_\alpha})(ip_4^{a_\beta})(ip_1^{b_1})^{P_1}(ip_3^{b_1})^{1-P_1}
G(p_1,p_2,p_3,p_4)
\cr\times
\lbrace
2\eta^{b_2m}(\eta^{na_3}p_3^{b_3}+\eta^{a_3b_3}p_3^n)
\Gamma({\vec{p_3}}{\vec{p_4}}+N_1+P_1-5)
\lbrack
{{\Gamma({\vec{p_1}}{\vec{p_3}}-N_1+5)}\over
{\Gamma(-{\vec{p_2}}{\vec{p_3}}+P_1)}}
\cr
+
{{\Gamma({\vec{p_1}}{\vec{p_3}}-N_1+3)}\over
{\Gamma(-{\vec{p_2}}{\vec{p_3}}+P_1-2)}}
\rbrack
\cr
+
\eta^{b_2m}(2\eta^{na_3}p_3^{b_3}-\eta^{a_3b_3}p_3^n)
\Gamma({\vec{p_3}}{\vec{p_4}}+N_1+P_1-6)
\lbrack
{{\Gamma({\vec{p_1}}{\vec{p_3}}-N_1+6)}\over
{\Gamma(-{\vec{p_2}}{\vec{p_3}}+P_1)}}
\cr
+
{{\Gamma({\vec{p_1}}{\vec{p_3}}-N_1+4)}\over
{\Gamma(-{\vec{p_2}}{\vec{p_3}}+P_1-2)}}
\rbrack\rbrace}}
and
\eqn\grav{\eqalign{
S_4^{(3)4-der}
=
24iA_m(p_1)A_n(p_2)H_{a_1a_2a_3}(p_3)H_{b_1b_2b_3}(p_4)
\sum_{M_1=0}^{1}\sum_{N_1=0}^1\sum_{P_1=0}^1
{{(-1)^{P_1}}}
\cr\times
(ip_3^m)^{M_1}(ip_4^m)^{1-M_1}
(ip_1^{a_1})^{N_1}(ip_4^{a_1})^{1-N_1}
(ip_1^{b_1})^{P_1}(ip_3^{b_1})^{1-P_1}
G(p_1,p_2,p_3,p_4)
\cr\times
\lbrace
2\eta^{a_2b_2}(\eta^{na_3}p_3^{b_3}+\eta^{a_3b_3}p_3^n)
\Gamma({\vec{p_3}}{\vec{p_4}}+N_1+P_1-6)
\lbrack
{{\Gamma({\vec{p_1}}{\vec{p_3}}-M_1-N_1+5)}\over
{\Gamma(-{\vec{p_2}}{\vec{p_3}}-M_1+P_1-1)}}
\cr
+
{{\Gamma({\vec{p_1}}{\vec{p_3}}-M_1-N_1+3)}\over
{\Gamma(-{\vec{p_2}}{\vec{p_3}}-M_1+P_1-3)}}
\rbrack
\cr
+
\eta^{a_2b_2}(2\eta^{na_3}p_3^{b_3}-\eta^{a_3b_3}p_3^n)
\Gamma({\vec{p_3}}{\vec{p_4}}+N_1+P_1-7)
\lbrack
{{\Gamma({\vec{p_1}}{\vec{p_3}}-M_1-N_1+6)}\over
{\Gamma(-{\vec{p_2}}{\vec{p_3}}-M_1+P_1-1)}}
\cr
+
{{\Gamma({\vec{p_1}}{\vec{p_3}}-M_1-N_1+4)}\over
{\Gamma(-{\vec{p_2}}{\vec{p_3}}-M_1+P_1-3)}}
\rbrack\rbrace}}
This concludes the computation of $S_4(1-1-3-3)$ 
and of $S(1-1-3-3)(p_1,...,p_4)$ in general. The overall 
 $4$-point amplitude $A(1-1-3-3)(p_1,...,p_4)$
 describing the $1-1-3-3$ quartic
interaction is obtained from $S(1-1-3-3)(p_1,...,p_4)$
by adding $A(1-1-3-3)(p_1,...,p_4)=S(1-1-3-3)(p_1,...,p_4)
+(p_3{\leftrightarrow}p_4)$, according to (34).

\centerline{\bf 4-point Amplitude and $1-1-3-3$ Quartic Interaction}

Now that our computation of the $1-1-3-3$ point amplitude is complete,
the concluding step is to deduce the related quartic interaction
from the structure of $A(1-1-3-3)$.
The momentum factors of $ip_J(J=1,...4)$ translate into
derivatives of the space-time fields $A^m,A^n,H^{a_1a_2a_3}$
and $H^{b_1b_2b_3}$ in the position space, while the common $f(p_1,p_2,p_3,p_4)$
factor reflects the nonlocality of the interaction.
In addition, each of the terms in the amplitude (34) contains 
 $\gamma$-function factor with the structure
\eqn\grav{\eqalign{\Xi(M_1,N_1,P_1)\sim
\Gamma({\vec{p_3}}{\vec{p_4}}-a(M_1,N_1,P_1))
\cr\times
{\lbrack}
{{\Gamma({\vec{p_1}}{\vec{p_3}}+b(M_1,N_1,P_1))}
\over{\Gamma(-{\vec{p_2}}{\vec{p_3}}-c(M_1,N_1,P_1))}}
+
{{\Gamma({\vec{p_1}}{\vec{p_3}}+b(M_1,N_1,P_1)-2)}
\over{\Gamma(-{\vec{p_2}}{\vec{p_3}}-c(M_1,N_1,P_1)-2)}}
\rbrack
}}
where $a,b$ and $c$ are the numbers appearing in
summations over $M_1,N_1$ and $P_1$.
The $\gamma$ function factor is of some subtlety.
While the numbers $a,b$ and $c$ differ from term to term
, it is easy to see that in general
$a>0,b>0$ and $b\geq{2}$  for each term in the amplitude.
For this reason, in the field theory limit 
${\vec{p_I}}{\vec{p_J}}\rightarrow{0}$ that we are interested in,
 $\Xi(M_1,N_1,P_1)$ generally includes singular part,
with simple poles in
${\vec{p_3}}{\vec{p_4}}$ and ${\vec{p_1}}{\vec{p_3}}$
( the latter only appear in terms with $b=2$), as well as the 
part regular in ${\vec{p_I}}{\vec{p_J}}$.
The singular part is actually related  to the flow of the
cubic part of the effective action, rather than to the genuine
quartic interaction we are looking for.
That is, the singular terms in the $\gamma$-function factors
 are related to two types of exchanges: 
the first is the  $s=1$ field exchange between two $s=3$ vertices,
while the second is the $s=3$ field exchange between $s=3$ and $s=1$
operators.
These are the exchanges that induce the RG flows on the worldsheet
for $s=1$ and $s=3$ fields,
resulting in the leading (cubic) order terms in the low energy effective
action.
 Schematically, the $\beta$-function 
of the $s=1$ field in the $s=3$ background
is given by 
$\beta_A\sim-{\Delta}A+{C}H^2$ where $C$ are the structure
constants appearing in the $1-3-3$ 3-point amplitude (expressed
in the position space). This particularly
leads to cubic terms of the type $\sim{CAH^2}$ 
in the low-energy effective action.
At the same time, the low-energy effective equations of motion
for the $s=3$ gauge field in the presence of $s=1$ background are given by, 
in the leading order,
$\beta_{H}{\sim}\Delta{H}-CAH=0$ which, if substituted into 
 cubic terms, lead to  ``nonlocal'' quartic terms of the type
$\sim{{{C^2A^2H^2}}\over{\Delta}}$ which structurally coincide
with the contribution of the singular part of the $\Gamma$-function
factor to the 4-point amplitude. To obtain the genuine 
quartic $1-1-3-3$ interaction from the $4$-point amplitude (34),
one has to subtract the singularities from each of the $\Gamma$-function
factors appearing in the expressions (42)-(63), similarly to the
 procedure explained
in ~{\tseytlin}.
The $\Gamma$-function factors with $b(M_1,N_1,P_1)>2$
can be expanded in ${\vec{p}}_I{\vec{p}}_J$ with the
leading order terms given by
\eqn\grav{\eqalign{\Xi(a(M_1,N_1,P_1),b(M_1,N_1,P_1),c(M_1,N_1,P_1))
\cr
=
\Gamma({\vec{p_3}}{\vec{p_4}}-a)
{\lbrack}
{{\Gamma({\vec{p_1}}{\vec{p_3}}+b)}
\over{\Gamma(-{\vec{p_2}}{\vec{p_3}}-c)}}
+
{{\Gamma({\vec{p_1}}{\vec{p_3}}+b-2)}
\over{\Gamma(-{\vec{p_2}}{\vec{p_3}}-c-2)}}
\rbrack
\cr\approx
{{(-1)^{a+c}(b-3)!(c-2)!{\vec{p}}_2{\vec{p}}_3}\over
{a!{\vec{p}}_3{\vec{p}}_4}}
\cr\times
\lbrace
1+(b-2)(b-1)(c-1)c
+({\vec{p}}_1{\vec{p}}_3)\lbrack
(b-2)(b-1)(c-1)cL(b-1)+L(b-3)\rbrack
\cr
+
({\vec{p}}_2{\vec{p}}_3)\lbrack
(b-2)(b-1)(c-1)cL(c)+L(c-2)\rbrack
\cr
-({\vec{p}}_3{\vec{p}}_4)\lbrack
(b-2)(b-1)(c-1)c+1\rbrack
+...}}
Then the related factors in the
quartic terms in the low energy effective action
are  given by
replacing each of the  $\Xi(a,b,c)$ factors in
the amplitude
according to
\eqn\grav{\eqalign{
\Xi(a,b,c)
\rightarrow
{\tilde{\Xi}}(a,b,c)
\cr=
\Xi(a(M_1,N_1,P_1),b(M_1,N_1,P_1),c(M_1,N_1,P_1))
-{{(-1)^{a+c}(b-3)!(c-2)!{\vec{p}}_2{\vec{p}}_3}\over
{a!{\vec{p}}_3{\vec{p}}_4}}
\cr\times
\lbrace
1+(b-2)(b-1)(c-1)c
+{\vec{p}}_1{\vec{p}}_3\lbrack
(b-2)(b-1)(c-1)cL(b-1)+L(b-3)\rbrack
\cr
+
{\vec{p}}_2{\vec{p}}_3\lbrack
(b-2)(b-1)(c-1)cL(c)+L(c-2)\rbrack
\cr
-{\vec{p}}_3{\vec{p}}_4\lbrack
(b-2)(b-1)(c-1)c+1\rbrack}}

where

\eqn\lowen{L(n)=\sum_{m=1}^n{1\over{m}}}

Each of ${\tilde{\Xi}}$
 contributes,
in the leading order in ${\vec{p_I}}{\vec{p_J}}$,
the factor given by

\eqn\grav{\eqalign{
{\tilde{\Xi}}(a,b,c)\approx
(-1)^{a+c+1}{\vec{p}}_2{\vec{p}}_3
{{(b-3)!(c-2)!}\over{a!}}L(a){\lbrack}1+(b-2)(b-1)(c-1)c\rbrack}}
Similarly, in the special case of $b=2$
the $\Xi\rightarrow{\tilde{\Xi}}$ replacement
for the quartic term in the low energy effective action 
is given by

\eqn\grav{\eqalign{\Xi(a,2,c)\rightarrow{\tilde{\Xi}}(a,2,c)
\cr
=\Xi(a,2,c)+
{{(-1)^{a+c+1}a!c!({\vec{p}}_2{\vec{p}}_3)}\over
{({\vec{p}}_3{\vec{p}}_4)}}
\lbrack{1}+{\vec{p}}_1{\vec{p}}_3+L(c){\vec{p}}_2{\vec{p}}_3\rbrack
\cr
+(-1)^{a+c}a!(c-2)!({1\over{{\vec{p}}_3{\vec{p}}_4}}
+
{1\over{{\vec{p}}_1{\vec{p}}_3}})(1+L(c-2)
{\vec{p}}_2{\vec{p}}_3)}}

with the leading order contribution

\eqn\grav{\eqalign{{\tilde{\Xi}}(a,2,c)
{\approx}(-1)^{a+c+1}a!c!L(a){{\vec{p}}_2{\vec{p}}_3}
+
(-1)^{a+c+1}a!(c-2)!(L(a)-2L(c-2))}}
to the amplitude.
This concludes the evaluation of the 
gauge-invariant $1-1-3-3$ quartic
interaction in the low energy effective action.

\centerline{\bf $1-1-5-5$ Structure. Conclusion and Discussion}

In this paper we have obtained gauge-invariant quartic interaction
of  massless higher spin fields in string theory approach.
Although we have concentrated on $1-1-3-3$ case, with the structure
of higher spin vertex operators
 basic properties of amplitudes discussed in this paper
 (such as nonlocality and derivative
structure of the kinematic part of the amplitude)
will also hold for more general $1-1-s-s$ cases.
The nonlocality structure of the 4-point amplitude
calculated in this paper is the consequence of specific the ghost
structure  of the vertex operators for the massless $s=3$ fields.
In particular, nonstandard ghost coupling of $s=3$ vertices
leads to two integrated vertices appearing in the 4-point amplitude
(contrary to one out of 4 integrated vertex in the standard
Veneziano case) producing the factor that diverges on-shell
but leads to nonlocalities in the $\beta$-function equations
(which essentially are the off-shell equations).
As for the local part of the $3-3-1-1$ interaction terms,
it is structurally reminiscent of the $3$-point $3-3-2$ amplitude
on the disc describing  $cubic$ gravitational couplings of  massless
spin $3$ field, that can be expressed in terms of linearized Weyl
tensor ~{\per}.
 In particular, the minimal number of derivatives in the kinematic part 
of $1-1-3-3$ is equal to $2\times{3}-2=4$, similar to
 the $3-3-2$ case. While it is known that
cubic $s-s-2$ gauge-invariant couplings always contain
a minimum of $2s-2$ space-time derivatives
~{ \per} it looks
plausible that similar derivative rule applies to the kinematic part
quartic $1-1-s-s$ couplings as well, as the disc 
$2-s-s$ amplidude (with two spin $s$ operators on the boundary and
the spin $2$ operator in the bulk) is structurally similar
to special limit ($p_3=p_4$) of the kinematic part of
the $1-1-s-s$ point  amplitude in open string theory
(where $p_3$ and $p_4$ are the momenta of the spin  
$s$ particles). 
So at least string theory appears to predict that
the minimal derivative rule for the quartic $1-1-s-s$ couplings
should  be similar to the one established for the $2-s-s$ case. 
This  certainly is the case for $s=3$ and
it would be interesting to check if this rule also
works for spins higher than $3$. Conceptually,
 the calculation performed in this paper for the $1-1-3-3$ case,
should be quite similar for $1-1-s-s$ amplitudes with higher
values of $s$ as well.
In any case, the derivative/momentum structure of the amplitudes
is tightly controlled by the ghost structure of the vertex operators
and by the overall ghost number balance.
For example,the
$1-1-5-5$ amplitude $A(1-1-5-5)(p_1,...,p_4)$ is structurally
\eqn\grav{\eqalign{
A(1-1-5-5)(p_1,...,p_4)=S(1-1-5-5)(p_1,...,p_4)+(p_3{\leftrightarrow}p_4)
\cr
S(1-1-5-5)
=<V_{s=1}^{(-2)}V_{s=1}^{(-2)}(0):\Gamma^{-1}V_{s=1}^{(-2)}:(\infty)
A^{(0)}_{s=3}(z_1)A^{(0)}_{s=3}(z_2)>}}
where, as previously
$V_{s=1}^{(-2)}=ce^{-2\phi}\partial{X^n}e^{ipX}A_n(p)$
are unintegrated photon operators at picture $-2$,
$\Gamma^{-1}=-4ce^{\chi-2\phi}\partial\chi$
is the inverse picture changing, so the photon at picture $-3$
has the overall ghost structure $\sim{e^{\chi-4\phi}}$, while
\eqn\grav{\eqalign{
A^{(0)}_{s=3}(z){\sim}H_{a_1...a_5}\oint{dw}(z-w)^4{e^{2\phi}}
P^{(4)}_{2\phi-2\chi-\sigma}
\partial{X^{a_1}}
\partial{X^{a_2}}\partial{X^{a_3}}
\partial\psi^{a_4}\psi^{a_5}{e^{ipX}}(w)}}
Note that, although the full BRST-invariant expression for
spin $5$ operators contains, apart from
$A^{(0)}$ terms with ghost structures $\sim{ce^{\chi+\phi}}$
and $\sim\partial{c}{c}e^{2\chi}$, the ghost balance condition
only allows the 
contributions from $A^{(0)}$-part with the ghost structure
$\sim{e^{2\phi}}$ (provided, of course, that the photons are chosen
at pictures $-2$ and $-3$). Again, we see that, first of all,
the amplitude contains a double worldsheet integration
(as in the $1-1-3-3$ case)
leading to the nonlocality of the interaction.
While the computation  of the matter/kinematic part of this amplitude 
is relatively straightforward
and similar to the $1-1-3-3$ case described above,
 the evaluation of the ghost part of this amplitude 
is  quite tedious  due to lengthy operator products
of the ghost polynomials $P^{(4)}_{2\phi-2\chi-\sigma}$  with the ghost
exponents and between themselves. Below we present
 the expression for the $1-1-5-5$ amplitude up  to numerical coefficients
which can be fixed by explicit evaluation of these operator products.
Evaluating the correlator (71) and integrating in $z_1,z_2$ we get the answer
\eqn\grav{\eqalign{S_{1-1-5-5}=
A_m(p_1)A_n(p_2)H_{a_1a_2a_3}(p_3)H_{b_1b_2b_3}(p_4)
\eta^{a_4b_4}\eta^{a_5b_5}\eta^{a_3n}
\cr\times
\sum_{L=0}^4\sum_{Q=0}^{2L}\sum_{Q_1=0}^Q\sum_{Q_2=0}^{2L-Q}\sum_{R_1=0}^{4-Q_2}
\sum_{M_1=0}^1\sum_{N_1=0}^2\sum_{P_1=0}^3
{{(-1)^{P_1}}\over{N_1!(2-N_1)!P_1!(3-P_1)!}}
\alpha_{L,Q,Q_1,Q_2,R_1}
\cr\times
\prod_{\alpha=1}^{N_1}\prod_{\beta=N_1+1}^2
\prod_{\gamma=1}^{P_1}\prod_{\lambda=P_1+1}^2
(ip_1^{a_\alpha})(ip_4^{a_\beta})(ip_1^{b_\gamma})(ip_3^{b_\lambda})
(ip_3^m)^{M_1}(ip_3^m)^{1-M_1}
\cr\times
G(p_1,p_2,p_3,p_4)
{{\Gamma(9+Q_1-Q+R_1-M_1-N_1+{\vec{p}}_1{\vec{p}}_3)
\Gamma(2L-20+N_1+P_1+{\vec{p}}_3{\vec{p}}_4)}\over
{\Gamma(P_1-M_1+R_1+Q_1-Q-11-{\vec{p}}_2{\vec{p}}_3)}}
\cr
+ (m\leftrightarrow{n},N_1\leftrightarrow{P_1})\times(-1)^{N_1+P_1}}}
where 
$\alpha_{L,Q,Q_1,Q_2,R_1}$ are the numerical coefficients to be
extracted from the ghost OPEs.
The overall amplitude $A(1-1-5-5)(p_1,...,p_4)$
is again obtained from $S(1-1-5-5)(p_1,...,p_4)$
by adding $A(1-1-5-5)=S(1-1-5-5)+(p_3{\leftrightarrow}p_4)$
according to (71).
The kinematic part of this amplitude contains minimum number of 6 space-time
derivatives.
At the same time, all the $\Gamma$-functions
in the denominator of (73) are proportional to
$\sim({\vec{p}}_2{\vec{p}}_3)^{-1}$ in the field
theory limit, for all the
values of $M_1,P_1,Q_1,R_1$ and $Q$. For this reason, the local part of the
 amplitude (73) is of at least 8 powers in momentum,
 again in according to the $2s-2$-rule conjectured above.
One has  to check explicitly, however, whether this rule
holds in each separate case for different values of $s$.

\centerline{\bf Acknowledgements}

It is a great pleasure to acknowledge the hospitality
of Center for Quantum Space-Time (CQUeST) at Sogang University
in Seoul where significant part of the results presented in this
paper has been obtained.
In particular,
I would like to express my deep gratitude  to Chaiho Rim
for his invitation to visit CQUeST and to Bum-Hoon Lee and Chaiho Rim
for their kind hospitality during my stay.
I also would like to thank Eoin Colgain, 
Robert De Mello Koch, Kimyeong Lee, Jeong-Hyuck Park,
Chaiho Rim, Augusto Sagnotti and Per Sundell
for interesting and useful discussions.

\listrefs

\end